# The electrical resistance of spatially varied magnetic interface. The role of normal scattering

R.N. Gurzhi, A.N. Kalinenko, A.I. Kopeliovich, P.V. Pyshkin, and A.V. Yanovsky

*B. Verkin Institute for Low Temperature Physics and Engineering of the National Academy of Sciences of Ukraine*
*47 Lenin Ave., Kharkov 61103, Ukraine*
E-mail: kopeliovich@mail.ru

We investigate the diffusive electron transport in conductors with spatially inhomogeneous magnetic properties taking into account both impurity and normal scattering. It is found that the additional interface resistance that arises due to the magnetic inhomogeneity depends essentially on their spatial characteristics. The resistance is proportional to the spin flip time in the case when the magnetic properties of the conducting system vary smoothly enough along the sample. It can be used to direct experimental investigation of spin flip processes. In the opposite case, when magnetic characteristics are varied sharply, the additional resistance depends essentially on the difference of magnetic properties of the sides far from the interface region. The resistance increases as the frequency of the electron-electron scattering increases. We consider also two types of smooth interfaces: (i) between fully spin-polarized magnetics and usual magnetic (or non-magnetic) conductors, and (ii) between two fully oppositely polarized magnetic conductors. It is shown that the interface resistance is very sensitive to appearing of the fully spin-polarized state under the applied external field.

PACS: 72.25.Mk Spin transport through interfaces;
73.40.Cg Contact resistance, contact potential.

Keywords: diffusive electron transport, spin flip processes, magnetic characteristics.

## 1. Introduction

The well-known and highly-applied giant magnetoresistance effect [1] is one of the effects which arise at contact between conductors with different magnetic properties. Really, the interface between two fully opposite polarized ferromagnetic conductors is an opaque obstacle for carriers as their spin polarizations are specified rigidly by the magnetization of the corresponding regions. The lesser effect arises at contact of a fully polarized magnetic conductor with a non-magnetic conductor as it was discussed in Ref. 2. Inserting a non-magnetic conductor between fully polarized magnetic sides (see Fig. 1) double its resistance (when the length of the non-magnetic part is less than the spin flip relaxation length $\lambda$). The reason is that one of the spin channels is cut off due to full polarization of the magnetic sides. Note, one may detect a non-equilibrium spin-polarization that exists in the magnetically inhomogeneous circuit [2], measuring its resistance. Really, spin polarization of a non-magnetic section disappears under demagnetization of the magnetic, and the interface resistance disappears too.

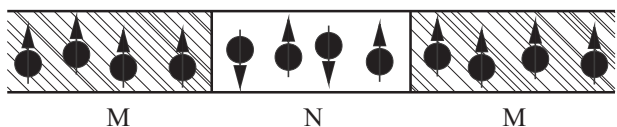

*Fig. 1.* M–N–M contact. M — fully polarized magnetic conductor, N — non-magnetic conductor.

Naturally, a contact between different magnetics is a source of the additional resistance. Spin-accumulation effects were investigated in the presence of the spin-dependent scattering early [3–6] in magnetic layered structures. These effects cause the interfacial resistance but electron-electron scattering, which conserves the total momentum of the system of interacting particles, is not to be used.

As it was firstly demonstrated in Ref. 7, the electron-electron scattering increases the interfacial resistance essentially. The physical reason is the mutual friction between "spin-up" and "spin-down" electrons. Let's assume that electric current flows from a non-magnetic conductor into a fully polarized magnetic region where all electrons are fully spin-polarized ("spin-up", for example, see Fig. 2). Then, "spin-down" electrons have no possibility to pass into the magnetic conductor and they are accumulated near the interface. In other words, a non-equilibrium additional spin density $\delta\rho_\downarrow$ arises in the non-magnetic conductor at the length-scale of the order of $\lambda$ from the interface. The value of $\delta\rho_\downarrow$ is determined by the following condition. The sum of the ohmic "spin-down" current and the diffusion "spin-down" current should be zero at the interface, i.e. the "drift" velocity of the "spin-down" component is zero. Thus, the crowd of "spin-down" unmovable electrons, which are accumulated near the interface, "slows down" the flow of "spin-up" electrons. In other words, the crowd of "spin-down" electrons plays the role of "effective" scattering impurities.

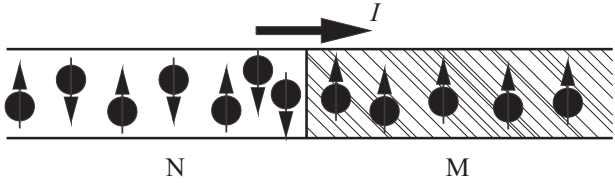

*Fig. 2.* The "crowd" effect: "spin-down" electrons, which are accumulated near the interface, "slow down" the flow of "spin-up" electrons.

As a result, the interfacial resistance increases indefinitely (and not just a twice) with increasing of the frequency of electron-electron collisions. Here we should note that electron-electron collisions are the "normal" collisions that conserve the total momentum of the system of interacting particles. So, they do not provide the resistance of the homogeneous parts of the electrical circuit but they give the essential contribution into the interface resistance. The contribution is proportional to the frequency of electron-electron collisions. Thus, the relative contribution of the interface resistance into the total circuit resistance increases with the electron-electron frequency increase. Clearly, any normal collisions will play the same role (e.g., electron-phonon collisions, when phonons are tightly coupled to the electron system). Note, that the role of electron-electron collisions was discussed in [7] for a particular case when uniform magnetic conductors contact with non-magnetic conductors. Below we analyze the problem for the general case of non-homogeneous conductors.

In this paper we propose some new possibilities for investigation of the role of a non-equilibrium spin density for electron transport properties. As we demonstrate below, a number of effects arises when electron spectrum is varied smoothly in space. Thus, a such type conductors are quite perspective for direct experimental research of non-equilibrium spin density effects. Firstly, one can exclude the influence of electrical contact barriers. Secondly, variations of magnetic properties can be easy induced by applying spatially inhomogeneous gate voltage in the case of a two-dimensional electrons in heterostructures (see Ref. 8), by variation of an external magnetic field or by space-dependent distributions of doping impurities. Note, we suppose a collinear magnetization of an inhomogeneous magnetic system. The additional source of spin mixing due to non-collinear spins has not being included into our consideration. The resistance of a smooth interface, that is a magnetic layer varied in spatial like a domain wall behavior, was calculated in [9].

The paper is organized as follows. The general equations for the resistance of a magnetically inhomogeneous conductor are derived in Sec. 2. The resistance is given in terms of the entropy production at electron diffusion and spin-flip processes. The case of relatively frequent spin-flip scattering is given in Sec. 3. Here we discussed the role of electron-electron collisions and their interplay with spin-flip scattering. The case of relatively rare spin-flip scattering can be found in Sec. 4. Electron transport through smooth almost fully spin-polarized interfaces is discussed in Sec. 5. A detailed derivation of the expressions of Sec. 4 is given in the Appendix.

In this paper we consider the diffusion transport regime and apply the modified set of equations were derived by Flensberg (see Ref. 10). We do not consider the influence of current flow on the electron spectrum, i.e. we do not suppose spin-torque effects [11] assuming that electric current is weak enough.

## 2. The electrical resistance of a smooth magnetic interface. The general approach

The role of electron-electron collisions for the interface resistance was demonstrated in Ref. 7 for the case of the zero-length interface between a non-magnetic conductor and a fully polarized magnetic material. Below we derive some general formulas for the electrical resistance of a conductor with an arbitrary spatial variation of the electron spectrum.

Let us rewrite the set of equations (1a), (1b) given in Ref. 10 in the following "vector" form for our one-dimensional task

$$\boldsymbol{\mu}' = -\hat{\beta}\mathbf{j}, \qquad (1)$$

$$\mathbf{j}' = -f(\boldsymbol{\mu} \cdot \mathbf{a})\mathbf{a}, \qquad (2)$$

$$\boldsymbol{\mu} = (\mu_\uparrow, \mu_\downarrow), \mathbf{j} = (j_\uparrow, j_\downarrow), \mathbf{a} = (1,-1),$$
$$\beta_{\uparrow\uparrow} = e(\rho_{i\uparrow} + An_\uparrow^{-2}), \beta_{\downarrow\downarrow} = e(\rho_{i\downarrow} + An_\downarrow^{-2}),$$
$$\beta_{\uparrow\downarrow} = \beta_{\downarrow\uparrow} = -eA(n_\uparrow n_\downarrow)^{-1}, f = \frac{e\Pi_0}{\tau_{sf}}, \qquad (3)$$
$$\Pi_0^{-1} = \Pi_\uparrow^{-1} + \Pi_\downarrow^{-1}.$$

Here $\mu_\uparrow$, $\mu_\downarrow$, $j_\uparrow$, $j_\downarrow$ are the spin components of the electrochemical potential and the current density, correspondingly. A prime denotes differentiation with respect to the coordinate along the conductor, $x$; $n_\uparrow$, $n_\downarrow$ are densities of "spin-up" and "spin-down" components, correspondingly; $\rho_{i\uparrow}$, $\rho_{i\downarrow}$ are the corresponding resistivities due to the processes with momentum loss, $e$ is the electron charge. $\Pi_{\uparrow,\downarrow}$ are the spin-dependent densities of states on the Fermi surface, $\tau_{sf}$ is the spin-flip characteristic time. The coefficient $A$ is proportional to the frequency of normal collisions $\nu_N$. In the case of the electron-electron scattering we have (see Ref. 7)

$$A \approx e^{-2} m \nu_{ee} n_m, \quad n_m^{-1} = n_\uparrow^{-1} + n_\downarrow^{-1}, \qquad (4)$$

where $m$ is a current carriers mass, $\nu_{ee}$ is the frequency of electron-electron collisions.

Let us specify the problem of the resistance of the inhomogeneous conductor in the following way. Let $L$ is the characteristic scale of the inhomogeneous part of the conductor which is homogeneous at $|x|>L$. The "resistivities" of its sides, $\hat{\beta}_l$ at $x<L/2$ and $\hat{\beta}_r$ at $x>L/2$, may be different from one another. The resistance of the conductor section between points $x=M$, $x=-M$, $M \gg L$, obviously, is given by $R=[\mu(-M)-\mu(M)]/ejs$. Here, $s$ is the area of the conductor cross-section(consider it to be constant), $\mu$ is the electro-chemical potential which is the same for the both spin components far from the inhomogeneity region; $js$ is the total current across the conductor cross-section. Evidently, by reason of electrical charge conservation $js$ does not depend on the $x$-coordinate. It is convenient to write that current in the following form: $j = \mathbf{j} \cdot \mathbf{n}$, where $\mathbf{n} = (1,1)$ and conservation of $j$ is seen in equation (2) scalary multiplied by $\mathbf{n}$.

Let us demonstrate that there is a relation between the electrical resistance of a magnetically homogeneous conductor and the entropy production rate which is similar to that in the case of a homogeneous conductor [12]. After integration by parts $\mathbf{j}\hat{\beta}\mathbf{j}$ and taking into account Eqs. (1) and (2), we obtain the following

$$R = \frac{1}{ej^2 s} \int_{-M}^{M} \left[ \mathbf{j}\hat{\beta}\mathbf{j} + f(\mathbf{\mu} \cdot \mathbf{a})^2 \right] dx. \qquad (5)$$

As it is easy to check, the quadratic form $\mathbf{j}\hat{\beta}\mathbf{j}/e$ is an essentially positive. Consequently, $R$ is positive too. The first term in subintegral expression of Eq. (5) corresponds to the entropy production at Joule heating. The second one corresponds to the entropy production due to the spin-flip scattering.

Let magnetic characteristics are varied rather smooth with $x$. So, the local equilibrium between spin-up and spin down-components had time to be established, i.e. $\mathbf{\mu} = \mu_e = \mu_e \mathbf{n}$, where $\mu_e(x)$ is an equilibrium electrochemical potential. In this case, the "equilibrium" current, as it follows from Eq. (1), is given by $\mathbf{j}_e = -\mu'_e \hat{\beta}^{-1} \mathbf{n}$. As we consider the total current to be fixed, $\mathbf{j}_e \cdot \mathbf{n} = j$. So, we may express the derivative $\mu'_e$ through the total current density directly

$$\mu'_e = -\frac{j}{\beta_{nn}^{-1}}, \quad \beta_{nn}^{-1} = \mathbf{n}\hat{\beta}^{-1}\mathbf{n}. \qquad (6)$$

Thus, we the "equilibrium" resistance is given by

$$R_e = \frac{1}{es} \int_{-M}^{M} \frac{dx}{\beta_{nn}^{-1}}. \qquad (7)$$

On the other hand, integrating the quadratic form $\mathbf{j}_e \hat{\beta} \mathbf{j}_e$ in parts and taking into account Eq. (1) and $\mu_e \cdot \mathbf{j}'_e = \mu_e (\mathbf{n} \cdot \mathbf{j}_e)' = 0$ yields

$$R_e = \frac{1}{ej^2 s} \int_{-M}^{M} \mathbf{j}_e \hat{\beta} \mathbf{j}_e \, dx. \qquad (8)$$

Equations (7) and (8) are equivalent. It is easy to see that the mentioned equilibrium state could be established in the case when the diffusion spin-flip length $\lambda$ is much less than the characteristic length-scale of the inhomogeneous interface region, i.e. $\lambda \ll L$. Here $\lambda \approx v\sqrt{\tau \tau_{sf}}$, $v$ is the carriers velocity, $\tau$ is the relaxation time which corresponds to scattering processes that change the momentum of an electron essentially (either with respect to the normal collisions or to the collisions that do not conserve the quasi-momentum of the electron system).

Let us define the addition to the total resistance, $\Delta R = R - R_e$, which arises due to the non-equilibrium spin density. The addition is a direct analog of the interface resistance between two homogeneous magnetic conductors. From Eqs. (5) and (8) we get

$$\Delta R = \frac{1}{ej^2 s} \int_{-\infty}^{\infty} \left[ \Delta\mathbf{j}\hat{\beta}\Delta\mathbf{j} + f(\mathbf{\mu} \cdot \mathbf{a})^2 \right] dx, \qquad (9)$$

$$\Delta\mathbf{j} = \mathbf{j} - \mathbf{j}_e.$$

Here we take into account that $\mathbf{j}_e \hat{\beta} \Delta\mathbf{j} = \Delta\mathbf{j}\hat{\beta}\mathbf{j}_e = -\Delta\mathbf{j} \cdot \mathbf{n}\mu_{e'} = 0$. The first equality is due to the fact that operator $\hat{\beta}$ is a self-adjoint operator. The second one is valid because $\Delta\mathbf{j}$ is an antisymmetric vector as to respect of the spin components: $\Delta\mathbf{j} = \Delta j \mathbf{a}$, $\Delta\mathbf{j} \cdot \mathbf{n} = 0$ when the total current density $j$ is fixed. We extend integration in Eq. (9) until infinite limits because $\Delta\mathbf{j}$ and $\mathbf{\mu} \cdot \mathbf{a}$ are vanishing in the homogeneous sides. Note, that $\Delta R$ is positive as it follows from Eq. (9). In other words, the total resistance increases because a non-equilibrium spin density arises due to the magnetic inhomogeneity.

Here we should note that diffusion current $\Delta j$ arose due to the non-equilibrium spin density. Let us define the diffusion coefficient for the non-equilibrium spin density. From Eq. (1) we get that $\mu' \cdot \mathbf{a} = -\beta_{aa}\Delta j$. Taking into account the well-known relations between the electrochemical potentials and electron densities, $\delta n_{\uparrow,\downarrow} = \Pi_{\uparrow,\downarrow} \mu_{\uparrow,\downarrow}$, and the condition of the electric neutrality, $\delta n_\downarrow = -\delta n_\uparrow$, we obtain the diffusion coefficient

$$D = \frac{1}{e\Pi_0 \beta_{aa}}, \beta_{aa} = e\left[ \rho_{i\uparrow} + \rho_{i\downarrow} + A\left(n_\uparrow^{-1} + n_\downarrow^{-1}\right)^2 \right].$$

Diffusive spin relaxation length is the following

$$\lambda = \sqrt{D\tau_{sf}} = \frac{1}{\sqrt{f\beta_{aa}}}. \qquad (10)$$

Equations (7) and (9) turn out to be more convenient for calculations of the resistance as to compare with the direct calculations of the electrochemical potential with the needed accuracy. Below we investigate theoretically the "non-equilibrium" addition to the resistance $\Delta R$ mainly. The reason is that one can separate the contributions of $R_e$ and $\Delta R$ into the total resistance, $R$, as they have different temperature and interface length-scale dependences. Moreover, in certain cases $\Delta R$ gives the main contribution.

### 3. A case of strong spin-flip scattering

At $\lambda/L \ll 1$, the spin equilibrium is established during the time when an electron passes diffusively the inhomogeneous interface region. Solving Eqs. (7) and (3) yields the "equilibrium" resistance

$$R_e = \frac{1}{se} \int_{-M}^{M} \frac{\rho_{i\uparrow}\rho_{i\downarrow} + A(\rho_{i\uparrow}n_{\downarrow}^{-2} + \rho_{i\downarrow}n_{\uparrow}^{-2})}{\rho_{i\uparrow} + \rho_{i\downarrow} + A(n_{\uparrow}^{-1} + n_{\downarrow}^{-1})^2} dx. \qquad (11)$$

This result demonstrates the following well-known fact [13]. The electron-electron scattering increases with temperature increasing and resistivity increases too. However, this transition does not change the order of magnitude of the resistivity while its value increases. Physically, the electron distribution transforms from the impurity formed one to the drift distribution which is typical for the strong electron-electron scattering. In the case, which is described by Flensberg's approximation [10], the impurity scattering forms a non-drift distribution, i.e. spin components have different drift velocities if $\rho_{i\uparrow}n_{\uparrow} \neq \rho_{i\downarrow}n_{\downarrow}$. Thus, in two opposite limiting cases the resistivity is given by different formulas in subintegral function of Eq. (11): $(\rho_{i\uparrow}\rho_{i\downarrow})/(\rho_{i\uparrow} + \rho_{i\downarrow})$ and $(\rho_{i\uparrow}n_{\downarrow}^{-2} + \rho_{i\downarrow}n_{\uparrow}^{-2})/(n_{\uparrow}^{-1} + n_{\downarrow}^{-1})^2$, correspondingly. Therefore, one can observe this transition experimentally when the temperature increases. However, there is no transition if one of the spin component is fully depleted (e.g., $n_\downarrow \to 0$) by applying of an external magnetic field or by electrical gating. The last statement is valid within the Flensberg's model which assumes that: (I) the electron spectrum is isotropic and (II) there are no groups of current carriers with different characteristics but with the same spin polarization. There is no temperature dependence of the resistance of a non-magnetic conductor within this model too. This is the consequence of spin degeneracy ($\rho_{i\uparrow}n_{\uparrow} = \rho_{i\downarrow}n_{\downarrow}$).

Now, let us calculate the contribution of the non-equilibrium spin density into the resistance, i.e. $\Delta R$. From Eq. (2) we get the antisymmetrical part of the electrochemical potential, $\mu \cdot a$, which is due the non-equilibrium spin density. To a first approximation, we put the "equilibrium" current in the right-hand-side of Eq. (2)

$$\mu \cdot a = -\frac{1}{2f}(j_e \cdot a)'. \qquad (12)$$

Here we take into account that $a^2 = 2$. As vector $\Delta j$, is antisymmetrical, Eq. (1) yields

$$\Delta j = -\frac{(\mu \cdot a)'}{\beta_{aa}}. \qquad (13)$$

Here and below matrix elements of operators in the basis of $a$ and $n$ are defined in the same way as in Eq. (6). It follows from Eqs. (12) and (13) that $\Delta j \ll j$ at $\sqrt{D\tau_{sf}} \ll L$. Thus, the approximation method used above for solving Eqs. (1) and (2) is verified when $\lambda \ll L$. As it follows from Eq. (13), within the same accuracy, the first term in the quadratic brackets in Eq. (9) is vanishing as to compared with the second one

$$\Delta j \hat{\beta} \Delta j = \frac{(\mu \cdot a)'^2}{\beta_{aa}} \ll f(\mu \cdot a)^2.$$

Therefore, from Eq. (9) we have

$$\Delta R = \frac{1}{4ej^2 s} \int_{-\infty}^{\infty} \frac{(j_e \cdot a)'^2}{f} dx,$$

where the equilibrium current $j_e$ can be found from Eqs. (1) and (6)

$$j_e = j \frac{\hat{\beta}^{-1} n}{\beta_{nn}^{-1}}. \qquad (14)$$

Finally, taking into account $\beta_{an}^{-1}/\beta_{nn}^{-1} = -\beta_{an}/\beta_{aa}$ we get the non-equilibrium spin density contribution to the resistance

$$\Delta R = \frac{1}{4es} \int_{-\infty}^{\infty} \frac{1}{f} \left(\frac{\beta_{an}}{\beta_{aa}}\right)'^2 dx =$$

$$= \frac{1}{4e^2 s} \int_{-\infty}^{\infty} \frac{\tau_{sf}}{\Pi_0} \left(\frac{\rho_{i\uparrow} - \rho_{i\downarrow} + A(n_{\uparrow}^{-2} - n_{\downarrow}^{-2})}{\rho_{i\uparrow} + \rho_{i\downarrow} + A(n_{\uparrow}^{-1} + n_{\downarrow}^{-1})^2}\right)'^2 dx. \qquad (15)$$

Note, unlike the "equilibrium" resistance given by Eq. (11), the non-equilibrium spin density contribution to the resistance, $\Delta R$, does not tend to zero at $\rho_{i\uparrow,\downarrow} \to 0$ (if $n_\uparrow \neq n_\downarrow$). The physical reason is the same that was described in Ref. 7: while normal collisions give no contribution into the resistance of homogeneous conductors, they cause the "crowd effect" in inhomogeneous system and, thus, increase the resistance. (Here we should note, that Eq. (11) corresponds to the approximation when the conductor is locally homogeneous).

Thus, Eq. (15) gives the main contribution to the resistance of the sample for the case when normal collisions dominate over other scattering processes: $\rho_i \ll An^{-2}$. (Here we neglected by the "side resistance" assuming that the length of the sample be of the order of the length-scale of the inhomogeneous interface region: $M \approx L$). Then we rewrite Eq. (15) in the following simple form

$$R = \frac{1}{4e^2 s} \int_{-\infty}^{\infty} \frac{\tau_{sf}}{\Pi_0} \left( \frac{n_\uparrow - n_\downarrow}{n_\uparrow + n_\downarrow} \right)'^2 dx. \qquad (16)$$

Summarize, Eq. (15) and, especially, Eq. (16) open a way for investigation of spin-flip processes by measuring the resistance of a conductor which magnetic properties are varied smoothly enough.

### 4. A case of weak spin flip scattering

In the case $L \ll \lambda$, we may neglect by the spin flip scattering inside the interface region within the main approximation. Then, the interface plays the role of a spin non-equilibrium density generator while spin relaxation occurs in the sides, out of interface. In other words, we may suppose the current density in the interface region, $\mathbf{j}_c$, is constant. Correspondingly, the electro-chemical potential is given by

$$\boldsymbol{\mu} = \boldsymbol{\mu}_c - \hat{B}(x)\mathbf{j}_c, \quad \hat{B}(x) = \int_0^x \hat{\beta}(x') \, dx', \qquad (17)$$

where $\boldsymbol{\mu}_c$ is a constant vector. Solutions of Eqs. (1) and (2) in homogeneous sides around the interface are the following

$$\Delta \mathbf{j} = c \mathbf{a} \, e^{\pm x/\lambda}, \quad \boldsymbol{\mu} = (\mu - bx)\mathbf{n} \mp \lambda c \hat{\beta} \mathbf{a} \, e^{\pm x/\lambda},$$

$$b_{l,r} = \frac{j}{\beta_{nnl,r}^{-1}}. \qquad (18)$$

Here index "*l*" and sign "+" corresponds to the left side, index "*r*" and sign "–" to the right side, respectively; $c_{l,r}$, $\mu_{l,r}$ are arbitrary constants that are defined below, the diffusive spin relaxation length $\lambda_{l,r}$ is given by Eq. (10).

One may match functions $\boldsymbol{\mu}$ in the interface region with its side asymptotes and find out the resistance. This seemingly simple task turns out to be relatively difficult. It is much easier to keep an accuracy of approximation applying Eq. (9). In the main approximation on the small parameter $L/\lambda$ we have to match currents

$$\mathbf{j}_c = c_l \mathbf{a} + \mathbf{j}_{el} = c_r \mathbf{a} + \mathbf{j}_{er}. \qquad (19)$$

Thus, we obtain from Eq. (9)

$$\Delta R e j^2 s = c_l^2 \lambda_l \beta_{aal} + c_r^2 \lambda_r \beta_{aar} + P. \qquad (20)$$

The first and second terms in Eq. (20) are the contributions of the boundary regions which are out of the interface. The length-scales of these regions are of the order of $\lambda$. To calculate these terms we used $f(\boldsymbol{\mu} \cdot \mathbf{a})^2 = f^{-1}(\mathbf{j}')^2$ (as it follows from Eq. (2)) and Eq. (10). The third term in Eq. (20) is the contribution of the interface region into the integral (9)

$$P = \int_{-\infty}^{\infty} \left[ \beta_{aa} (\mathbf{j}_c - \mathbf{j}_e)_a^2 - \beta_{aasd} c_{sd}^2 \right] dx. \qquad (21)$$

Here subscripts "*a*" at vectors denote their components along the ort **a**: $\mathbf{a} \cdot \mathbf{c} = 2c_a$; index "*sd*" marks the value of the given function in the corresponding boundary

$$\beta_{aasd} = \beta_{aar} \text{ at } x > 0, \beta_{aasd} = \beta_{aal} \text{ at } x < 0. \qquad (22)$$

We have subtracted $\beta_{aasd} c_{sd}^2$ from the integrand in Eq. (21) to keep integral convergence. In this way we separate the intrinsic and extrinsic (as to the interface region) contributions in Eq. (20).

In Appendix we give equations for $c_{l,r}$, $j_{ca}$, see Eqs. (A.2), (A.3). Thus, to obtain the non-equilibrium addition to the resistance in the case of the "short" interface ($\lambda \gg L$) it is enough to put Eq. (A.2) and (A.3) into the Eqs. (20) and (21).

Let us suppose that $\beta_{aa} \ll \beta_{aasd} \lambda / L$, i.e. there is no leap of the electrical resistance in the interface region (for the both spin components). Then, from Eqs. (20), (21) we obtain (see Appendix)

$$\Delta R = \Delta R_j + \Delta R_t, \qquad (23)$$

$$\Delta R_j = \frac{\lambda_l \lambda_r \beta_{aal} \beta_{aar} \left[ (\beta_{an}/\beta_{aa})_r - (\beta_{an}/\beta_{aa})_l \right]^2}{4es(\lambda_l \beta_{aal} + \lambda_r \beta_{aar})}, \qquad (24)$$

$$\Delta R_t = \frac{1}{4es} \int \beta_{aa} \left[ \frac{\beta_{ansd}}{\beta_{aasd}} - \frac{\beta_{an}}{\beta_{aa}} \right]^2 dx. \qquad (25)$$

Here $\Delta R_t$ is the direct contribution of the interface region. Generally speaking, $\beta_{sd}$ depends on the origin of the coordinate. But it is a technique only that allows us to keep the convergence of the integral in Eq. (25). Thus, the choice of the origin does not affect the result in the main approximation on the small parameter $L/\lambda$. The reason is that $\Delta R_t$ is not small as to compare with $\Delta R_j$ when $\hat{\beta}_r$ and $\hat{\beta}_l$ are close to each other.

Resistance $\Delta R_j$ (see Eq. (23)) arises due to the difference in magnetic properties between interface sides. This contribution was calculated [3,4] for the case when electron-electron scattering is neglecting. Note, Eq. (24) is valid even if $L > \lambda$ in the case, when the interface region has a complex structure and includes a number of narrow sub-interfaces which length $L_{tr}$ is much less than the diffusion spin flip length $\lambda$: $L_{tr} \ll \lambda$. It is the case, in Eqs. (15) and (16) one should exclude from integration that regions where derivative with respect to *x*-coordinate is diverged. Then, like to the general case of short interfaces, the con-tribution of an subinterface is given by Eq. (24) where indexes "*l*" and "*r*" correspond to the sub-interface boundaries. (Note, the given approach is valid when distances between interfaces exceed $2\lambda$, see [5].)

As it follows from Eqs. (23)–(25), the electron-electron scattering gives contribution into the interface resistance ($\beta_{aa}, \beta_{an} \propto \nu_i + \nu_{ee}$, see Eq. (3), where $\nu_i$ and $\nu_{ee}$ are the electron-impurity and the electron-electron frequencies of scattering). Moreover, the relative contribution of the spin non-equilibrium density, $\Delta R / R_e$, rises with the electron-electron scattering increasing as it gives no contribution into the "equilibrium" resistance $R_e$. Thus, the "crowd effect" [7] exists in the case of a smooth interface at an arbitrary relation between $L$ and $\lambda$ (the case when $L \gg \lambda$ was discussed in Sec. 3). In the case of different magnetic sides, a difference between drift velocities of the spin components disappears at the length-scale of the order of $\lambda$ deep into the side. There is a strong mutual friction between "spin-up" and "spin-down" components that leads to appearing of the resistance. In the case of the same magnetic sides, the electron-electron scattering equilibrates the difference in drift velocities mainly over a length of the interface.

Let us analyze the temperature dependence of the resistance. Taking into account that the diffusion length decreases at temperature increase (see Eq. (10)), we get $\Delta R \propto \sqrt{\nu_{ee}}$, when the electron–electron scattering dominates over the electron-impurity scattering and sides have different magnetic characteristics. On the other hand, $\Delta R \propto \nu_{ee}$ when magnetic characteristics of the sides are identical to each other. $\Delta R$ increases with $\nu_{ee}$ until $L \ll \lambda$. In the opposite limit case, Eqs. (15) and (16) are valid.

Note, the experimental results like that were obtained in Ref. 2 (fully-polarized magnetic sides which are separated by a non-magnetic insertion) can be described by equation (25) for the case of continuous transition along coordinate between magnetic materials and the non-magnetic insertion ($\beta_{ansd} / \beta_{aasd} = 1$). However, when that contact is "sharp", Eq. (25) gives the same result as Ref. 7. Here we have to stress on the difference between this result and results given below in Sec. 5. In contrast to the results of Sec. 5, the result discussed above is not so sensitive to appearing of the fully spin-polarized state.

In conclusion of this Section let us demonstrate the dependence of the additional resistance on the length of the interface, $L$, for the case when magnetic characteristics are varied weakly. Let "electroconductivity tensor" be written as $\hat{\alpha} \equiv \hat{\beta}^{-1} = \hat{\alpha}_0 + \delta\hat{\alpha}$, where $\hat{\alpha}_0$ does not depend on the coordinate $x$ and $\delta\hat{\alpha} \ll \hat{\alpha}_0$. It is easy to get solutions of Eqs. (1), (2) which are approximately valid for any relation between $L$ and $\lambda$. Let, our conductor is a non-magnetic in the main approximation, i.e. $\alpha_{0an} = \alpha_{0na} = 0$. Then we obtain the first order correction to the $\mu_a$ on the small parameter $\delta\hat{\alpha}$

$$\delta\mu_a(x) = -\frac{\lambda j}{2\alpha_{0nn}\alpha_{0aa}} \int e^{-|x-x'|/\lambda} \delta\alpha'_{an}(x')\, dx'. \quad (26)$$

In order to find $\delta\mu_n$ we have to put zero the symmetrical correction to the total current with the accuracy to the second order terms: $\alpha_{0nn}\delta\mu'_n + \delta\alpha_{an}\delta\mu'_a + \delta\alpha_{nn}\mu'_{n0} = 0$, where $\mu'_{n0} = -j/\alpha_{0nn}$. As a result, the resistance $\delta R = \delta R_e + \Delta R$ is given by

$$\delta R_e = -\frac{1}{\alpha^2_{0nn}} \int \delta\alpha_{nn}(x)\, dx,$$

$$\Delta R = \frac{\lambda}{2\alpha^2_{0nn}\alpha_{0aa}} \int e^{-|x-x'|/\lambda} \delta\alpha'_{an}(x)\delta\alpha'_{an}(x')\, dxdx'. \quad (27)$$

In Fig. 3 we plotted the additional resistance $\Delta R$ as a function of the length $L$ for the cases of identical (*a*) and different (*b*) sides of the interface. Note, $\Delta R \propto L^{-1}$ at $L \gg \lambda$ in both cases. Magnetic properties of the interface were modelled by the following way:

a) $n_{\uparrow,\downarrow} = n_0(1 \mp 0.5\, e^{-x^2/L^2})$,

b) $n_{\uparrow,\downarrow} = n_0(1 \mp 0.5\pi^{-1/2} \int_{-\infty}^{x} e^{-y^2/L^2}\, dy)$.

Solid curves were calculated from the solutions given by Eq. (27). The dashed line (Fig. 3,*a*) represents the depen-dence $\Delta R(L)$ in the limit $L \ll \lambda$. It was calculated from the solution given by Eq. (25). Dotted lines (Fig. 3,*a* and Fig. 3,*b*) correspond to the limit $L \gg \lambda$. These curves were calculated for the given model from Eq. (15). The result for the "zero-length" interface (shown by the triangle in Fig. 3,*b*) was calculated from Eq. (24). The additional equilibrium resistance $\delta R_e$ is, obviously, proportional to the $L$ for the case of identical sides and it does not depend on the electron-electron scattering. Thus, it is almost temperature independent and one could eliminate this contribution by the special choice of $\delta \hat{\alpha}$: it should be "antisymmetric" either as to the coordinate or as to the spin projection (at $\delta \alpha_{\downarrow} = -\delta \alpha_{\uparrow}, \delta \alpha_{nn} = 0$).

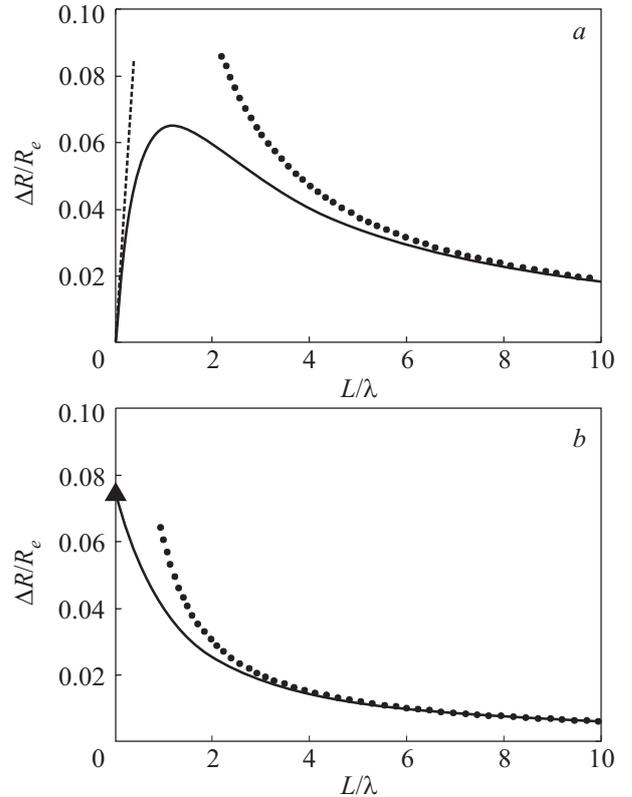

*Fig. 3*. The relative interfacial resistance $\Delta R / R_e$ as a function of the normalized length of the interface $L / \lambda$ for the cases of identical (*a*) and different (*b*) sides of the interface. The length of the sample is assumed as $10\lambda$ and $\nu_{ee} = 5\nu_i$

## 5. "Spin-stop" interfaces

In this section we discuss the case when the interface forms an essential barrier for electron transport (at least for one of the spin components). In the case when $L\beta_{aa}$ (which is proportional to the sum of interface resistivities for both spin channels) becomes of the order of the "boundary resistances" out of the interface, $\lambda \beta_{aasd}$ (which are due to the spin non-equilibrium densities), we can't neglect by the terms of the order of $\Delta B$ in Eqs. (A.2), (A.3) in spite of $L/\lambda \ll 1$.

In the case when sides around interface are identical to each other and $L\beta_{aa} \gg \lambda \beta_{aasd}$, we obtain from Eqs. (20), (21) and Eqs. (A.2), (A.3)

$$\Delta R = \frac{\lambda_s \beta_{aasd}}{2es} \left[ \frac{\beta_{ansd}}{\beta_{aasd}} - \frac{\Delta B_{an}}{\Delta B_{aa}} \right]^2 +$$
$$+ \frac{1}{4es} \int_{-\infty}^{\infty} (\beta_{aa} - \beta_{aasd}) \left[ \frac{\beta_{an}}{\beta_{aa}} - \frac{\Delta B_{an}}{\Delta B_{aa}} \right]^2 dx,$$
$$\Delta \hat{B} = \Delta \hat{B}(\infty) - \Delta \hat{B}(-\infty). \quad (28)$$

Here, the second term does not increase infinitely with $\beta_{aa}$ and $\beta_{an}$ increase. The reason is that the equality $\beta_{an}/\beta_{aa} = \Delta B_{an}/\Delta B_{aa}$ is valid with the good accuracy in that region where $\beta_{aa}$ and $\beta_{an}$ are large enough. That is why the second term in Eq. (28) is an essential only in that case, when the first term is vanishing (i.e. when the difference in quadratic brackets is neglecting). Thus, as follows from (A.1) and (28), the "leap" of resistivity causes the appearance of non-small spin non-equilibrium density out of the interface region at the length-scale of the order of $\lambda$ around it. It is the effect that gives the main contribution into $\Delta R$. Equation (28) gives the same result as corresponding equations in Ref. 7 for the case when fully polarized magnetic region ($n_{\uparrow} \to 0, \beta_{an}/\beta_{aa} \to \Delta B_{an}/\Delta B_{aa} \to 1$) is sided by non-magnetic conductors ($\beta_{ansd} = 0$).

Let us discuss a gated magnetic interface with controlled density of the spin components. Here one may achieve a fully polarized magnetic state by applying an external field to the gate thus depleting one of the spin components, $n_{\uparrow} \to 0$ (we call it as a "spin-stop" interface). As it follows from Eqs. (3), (4), the corresponding resistivity increases $\beta_{\uparrow\uparrow} \propto n_{\uparrow}^{-1}$ (in the two-dimensional case both impurity assisted resistance, $\beta_{i\uparrow} \propto \Pi_{\uparrow}/n_{\uparrow}$, and electron-electron contribution are proportional to $n_{\uparrow}^{-1}$). Consequently, $\beta_{aa}, \beta_{an} \propto n_{\uparrow}^{-1}$.

Thus, within the validity of Eq. (25) the contribution of the each point of the interface region into the $\Delta R$ rises as $n_{\uparrow}^{-1}$. The increase is limited by the value which is given by Eq. (28) when $L\beta_{aa}$ is of the order of $\lambda \beta_{aasd}$. The reason is that "spin-up" current can't enter into the fully "spin-down" polarized region when $L\beta_{aa} \gg \lambda \beta_{aasd}$, so it have to be converted into the "spin-down" component out that region. Meanwhile, at $L\beta_{aa} \ll \lambda \beta_{aasd}$ the region does not limit "spin-up" current flow.

Figure 4 demonstrates the increase of the interfacial resistance at depleting of the "spin-up" component due to the electrical gating in a two-dimensional magnetic conducting heterostructure ($n_{\uparrow} \to 0$ at $x = 0$). Here we assume that temperature is zero and use the following model for the electrical potential: $e\varphi/\varepsilon_{F\uparrow} = U_0 \exp(-x^2/L^2)$. Here $\varepsilon_{F\uparrow}$ is the Fermi energy for "spin-up" electrons and full depleting is achieved at $x = 0$. For the sake of specificity, let us assume that equilibrium spin densities for "spin-up" and "spin-down" electrons are related to each other as $n_\uparrow / n_\downarrow = 1/2$ far from the gate ($|x| \gg \lambda$). Then we may write the $x$-dependencies of the spin densities in the following way:

$$n_\uparrow = n_0(1 - U_0\, e^{-x^2/L^2}), \quad n_\downarrow = n_0(2 - U_0\, e^{-x^2/L^2}).$$

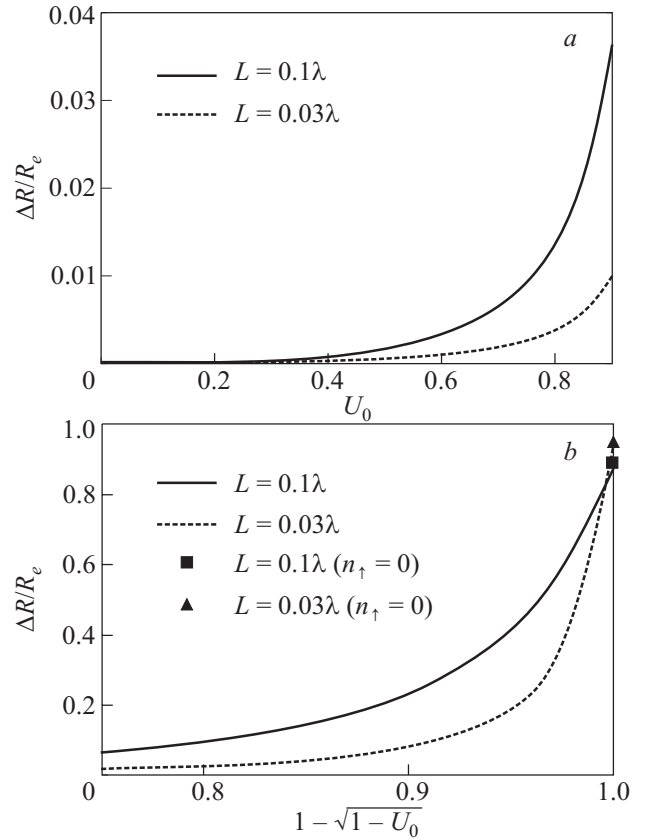

*Fig. 4.* The relative interfacial resistance as a function of the relative gate voltage $U_0 = e\varphi_{\max}/\varepsilon_{F\uparrow}$ (*a*). $\Delta R/R_e$ plotted as a function of $1 - \sqrt{1 - U_0}$ to reveal the characteristic features at depleting one of the spin components. The length of the sample is $2\lambda$ and assumed $\nu_{ee} = \nu_i$ (*b*).

Both solid and dotted curves were calculated from the solution given by Eq. (20). Note, at $U_0 \to 1$, these curves tends to the values (that marked by the triangle and square, correspondingly) which were calculated from Eq. (28) for the limit case of the zero-spin density for one of the spin components. It is easy to see, the resistance is very sensitive to formation of the fully spin-polarized state. Thus, it can serve as a marker of this state appearing.

In the opposite case, when the spatial transition to the fully spin-polarized magnetic is smooth enough ($L \ll \lambda$) the spin-flip scattering converts current with one spin polarization to a current of the opposite polarization with penetration current deep into the fully spin-polarized magnetic region. Thus, there is no large additional resistance of the interface. There are no divergences in Eqs. (11), (15) and (16) at $n_\uparrow \to 0$.

Within the frames of our approach we may calculate also the resistance of the smooth interface between two opposite fully spin-polarized magnetic sides (GMR contact). Let the left side is "spin-up" polarized ($n_\downarrow = 0$) and the right one is "spin-down" polarized ($n_\uparrow = 0$). Let us suppose also that there is an interface region between spin-polarized sides which length is $L_{\rm tr} \ll \lambda$ where "spin-up" and "spin-down" densities are not zero. Within the main approximation on the small parameter $L_{\rm tr}/\lambda$ the electrochemical potential $\mu_a$ does not depend on the coordinate $x$ as it has enough time to be balanced inside the interface region while spin-flip process occurs. Thus, within this approximation we get the following solution of Eq. (2)

$$j_a = -\mu_a \int_l^x f(x')\, dx' + j, \quad \mu_a = j\left[\int_l^r f(x')\, dx'\right]^{-1},$$

$$\mu_n' = -(\hat{\beta}\mathbf{j})_n \ll \mu_a / L_{\rm tr}. \qquad (29)$$

Here, we take into account the following boundary conditions: $j_\downarrow = 0$ at $x = x_l$, and $j_\uparrow = 0$ at $x = x_r$. As it follows from Eq. (1), $x$-dependent parts of the electrochemical potentials $\mu_{\uparrow,\downarrow}$, are of the order of $\mu_a(\lambda/L_{\rm tr})^2$. Really, while $\beta_{\downarrow\downarrow}$ increase rapidly near the left boundary, the product $\beta_{\downarrow\downarrow} j_\downarrow$ remains bounded above as $j_\downarrow$ tends to zero. Thus, our assumption on the $x$-independence of the $\mu_a$ is valid. Within the accuracy of the model we have take into account the second term only in the subintegral function in Eq. (9). Thus we get for the resistance of the interface

$$R_{\rm inter} = \left[es \int_l^r \frac{\Pi_0}{\tau_{sf}}\, dx\right]^{-1}. \qquad (30)$$

## 6. Summary

In summary, we have investigated the diffusive electron transport in conductors with spatially inhomogeneous magnetic properties taking into account both impurity and normal scattering. We have obtained the general equations for the electrical resistance of spatially inhomogeneous magnetic interfaces with collinear magnetization(see Sec. 2). The equations open an effective way to calculate the interfacial resistance. We found that spatial magnetic inhomogeneity causes the additional interfacial resistance which depends essentially on the spatial characteristics. If interfacial inhomogeneity is smooth enough, the spin non-equilibrium density causes the additional resistance the value of which is determined by spin-flip processes. It can be used for direct experimental investigation of spin flip processes (see Sec. 3).

The simplest relation between spin flip processes and the additional interfacial resistance arises when electron-electron scattering dominates over other scattering processes. In the case when the length-scale of an inhomogeneous interface region is short enough, we found separately both the contribution of the interface region into the resistance and the contribution of homogeneous sides (see Section 0.4). The interfacial resistance increases with the increase of the electron-electron scattering frequency. The reason is the mutual friction between "spin-up" and "spin-down" electron subsystems and the "crowd" effect.

We have found also the resistance of smooth "spin-stop" interface that means an essential barrier for electron transport of one of the spin components. We have demonstrated the sensitivity of the interfacial resistance to formation of a fully spin-polarized magnetic under the influence of applied external fields. We have demonstrated also that resistance measurements provide direct information on the frequency of spin-flip processes when both sides have antiparallel spin orientation. It's shown also in Sec. 3 that the formation of a fully spin-polarized magnetic shows itself in the temperature dependence of the "spin-equilibrium resistance".

The work was supported in part by NanoProgram of the NAS of Ukraine and NASU Grant F26-2.

### Appendix

To find constants $c_{l,r}$ in Eqs. (20), (21) we have to write the antisymmetrical part of the difference of the matching equations for the electro chemical potentials in the sides of the interface region

$$\lambda_r c_r \beta_{aar} + \lambda_l c_l \beta_{aal} + \Delta B_{aar} c_r + \Delta B_{aal} c_l + \\ + \mathbf{a}(\Delta \hat{B}_r \mathbf{j}_{er} + \Delta \hat{B}_l \mathbf{j}_{el}) = 0, \qquad (A.1)$$

where

$$\Delta \hat{B}_l = \int_{-\infty}^{0} \left[\hat{\beta}(x) - \hat{\beta}_l\right] dx, \quad \Delta \hat{B}_r = \int_{0}^{\infty} \left[\hat{\beta}(x) - \hat{\beta}_r\right] dx.$$

Equation (10) is written in the main approximation on the small parameter $L/\lambda$. Here we have taken into account $\mathbf{a}\hat{\beta}_{r,l}\mathbf{j}_{er,l} = -\mathbf{a}\cdot\mathbf{\mu}_{er,l} = 0$. Note that the form of equation does not depend on the choice of the *x*-coordinate of matching points out of the interface region. Really, expanding the exponent in Eq. (18) into the series near the interface region we obtain $\mathbf{\mu}\cdot\mathbf{a} = \mp\lambda c\beta_{aa}e^{\pm x/\lambda} \approx \mp\lambda c\beta_{aa} - xc\beta_{aa}$. At matching the second term will be cancelled by the *x*-dependent part of $\mathbf{\mu}$ from Eq. (17). For example, at matching in the right-hand side we get $-\mathbf{a}\hat{B}(x)c\mathbf{a} = -\Delta B_{aar}c_r - xc_r\beta_{aar}$. There is no need to keep next order terms on the parameter $L/\lambda$. Solving (A.1) together with (19) yields

$$c_{r,l} = -\frac{\pm\lambda_{l,r}\beta_{aal,r}(\mathbf{j}_{er} - \mathbf{j}_{el})_a + \mathbf{a}\Delta\hat{B}\mathbf{j}_{er,l}}{\lambda_r\beta_{aar} + \lambda_l\beta_{aal} + \Delta B_{aa}}, \quad \text{(A.2)}$$

$$\Delta\hat{B} = \Delta\hat{B}_r + \Delta\hat{B}_l.$$

Thus, we get for $j_{ca}$ from Eq. (19) the following

$$j_{ca} = \frac{2\lambda_r\beta_{aar}j_{ear} + 2\lambda_l\beta_{aal}j_{eal} - \Delta B_{an}j}{2(\lambda_r\beta_{aar} + \lambda_l\beta_{aal} + \Delta B_{aa})}, \quad \text{(A.3)}$$

$$j_{ea} = -j\frac{\beta_{na}}{\beta_{aa}}.$$

Here we take into account $j_{enr} = j_{enl} = j/2$.

Substituting Eqs. (A.2), (A.3) into Eqs. (20) and (21) give us a cumbersome equation for $\Delta R$. However, it can be simplified essentially when there is no "leap" of the resistance in the interface region, i.e. one may neglect by the $\Delta B$ in Eqs. (A.2), (A.3)) as to compare with $\lambda\beta_{aa}$. Then, the contribution of the interface region, $P$, is not vanishing as to compare with contribution of the regions of the diffusion spin non-equilibrium density, $\lambda_s\beta_{aasd}j^2$, in that case only when side conductivities of are close to one another: $(|j_{ear} - j_{eal}|/j)^2 \lesssim L/\lambda$. Then, one can neglect both by the difference between $j_{ear}$ and $j_{eal}$ and by the value $c_s \sim |j_{ear} - j_{eal}|/j$. As a result, after substituting Eqs. (A.2), (A.3) in Eqs. (20), (21) we get the formulas for the addition to the resistance, i.e. Eqs. (23)–(25).